# A Probabilistic Approach to Risk Mapping for Mt. Etna

Vena pearl boÑgolan[1], Rocco Rongo[2], Valeria Lupiano[2], Donato D'Ambrosio[3], William Spataro[3] and Giulio Iovine[4]

[1] Department of Computer Science, College of Engineering

University of the Philippines Diliman

bongolan@dcs.upd.edu.ph

[2] Department of Biology, Ecology and Earth Sciences, University of Calabria

[3] Department of Mathematics and Computer Science, University of Calabria

[4] CNR-IRPI, Cosenza, Italy

## ABSTRACT

We evaluate susceptibility to lava flows on Mt. Etna based on specially designed die-toss experiments using probabilities for type of activation, time and place culled from the volcano's 400-year recorded history and current studies on its known factures and fissures. The types of activations were forecast using a table of probabilities for events, typed by duration and volume of ejecta. Lengths of time were represented by the number of activations to expect within the time-frame, calculated assuming Poisson-distributed inter-arrival times for activations. Locations of future activations were forecast with a probability distribution function for activation probabilities. Most likely scenarios for risk and resulting topography were generated for Etna's next activation; the next 25, 50 and 100 years. Forecasts for areas most likely affected are in good agreement with previous risk studies made. Forecasts for risks of lava invasion, as well as future topographies might be a first. Threats to lifelines are also discussed.

KEY WORDS: probabilistic modelling, risk and susceptibility

## INTRODUCTION

Mount Etna is the most active volcano in Europe. It is composed by several nested strato-volcanoes, plus scattered eruptive centres, which opened through a basement of fissural basalts dating back to 500 ky BP, overlying Quaternary sediments covering the Maghrebian–Apennine Chain. The dominant type of activity is effusive. Structural discontinuities are not homogeneously distributed on the volcano, but concentrate into "sheaves" (see CRISCI, IOVINE, DI GREGORIO AND LUPIANO, 2008, and references therein).

In last decades, the vulnerability of the Etna area has increased exponentially, due to urbanization (DIBBEN, 2008). Therefore, a scenario of expected lava fields may allow for appropriate planning and civil defense actions. We emphasize here that lava flows come from vent activations or 'flank eruptions' and are different from crater eruptions. This latter is what is usually associated with 'volcanic eruptions', which are modelled differently.

The hazard induced by dangerous flow-type phenomena – e.g. lava flows, earth slows, debris flows, and debris avalanches – can be assessed by analyzing a proper set of simulations of hypothetical events (e.g. IOVINE, DI GREGORIO AND LUPIANO, 2003). Recently, D'AMBROSIO, RONGO, SPATARO, AVOLIO AND LUPIANO (2006), CRISCI, IOVINE, DI GREGORIO AND LUPIANO (2008) and CRISCI, AVOLIO, BEHNCKE, D'AMBROSIO, DI GREGORIO, LUPIANO, NERI, RONGO AND SPATARO (2010), TARQUINI & FAVALLI (2010), RONGO, AVOLIO, BEHNCKE, D'AMBROSIO, DI GREGORIO, LUPIANO, NERI, SPATARO and CRISCI, (2011), and CAPPELLO, VICARI, AND DEL NEGRO (2011) adopted a modelling approach for the evaluation of lava-flows hazard at Mount Etna (Italy). In their studies, simulated lava flows started from the nodes of regular grids of vents uniformly covering the study areas. A probability of occurrence could be assigned to each simulation, based on statistics of historical events: the spatial hazard was then obtained by simply considering the summation of the probabilities associated to the simulated flows affecting each point of the study area. In such examples, a huge number (over 100.000) of simulations had to be performed, thus requiring long-running computations for the assessment of the hazard.

## THE NON-UNIFORM GRID OF HYPOTHETICAL VENTS

As in CRISCI, AVOLIO, BEHNCKE, D'AMBROSIO, DI GREGORIO, LUPIANO, NERI, RONGO AND SPATARO (2010), a Probability Distribution Function (PDF) was considered to represent the zones with differing probabilities of eruptive vents opening. Specifically, the PDF proposed by LUPIANO (2011) was refined by the function kernel density estimation (Silverman, 1986) using in particular a Gaussian kernel, taking into account the historical distribution of lateral and eccentric vents, and the distribution of the main faults/structures on the volcano. Accordingly, the adopted grid of vents was made progressively finer at higher probabilities, in proportion to the probability of vent opening (see Fig. 1).
A non-uniform grid of hypothetical vents was then overlaid on the PDF, with the greatest density of vents in the red areas, half the density in the orange areas, a fourth of the density in the yellow areas, decreasing logarithmically to lowest density in the blue areas, where there is the least probability of opening of eruptive vents. This was done to fully exploit available knowledge of the topography, which is finer than what the previous uniform grid made use of.

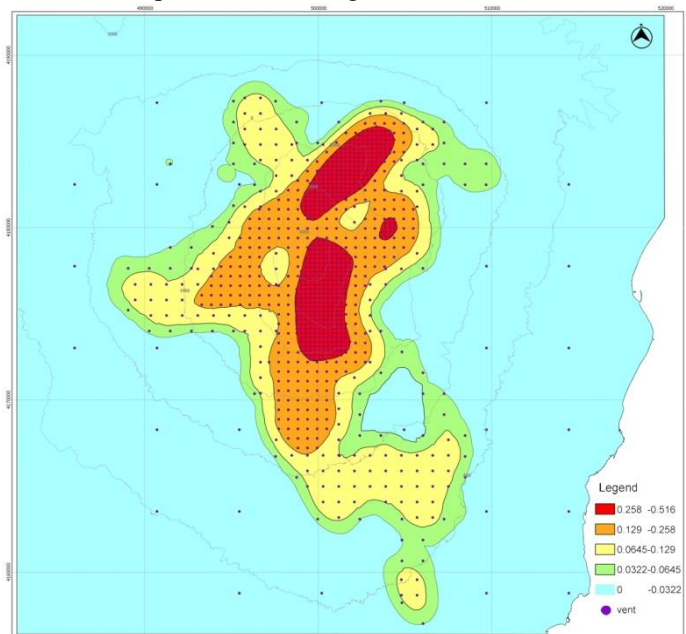

*Fig. 1* – Characterization of probabilities of activation (classes

of activation) of eruptive vents, and non-uniform grid distribution of 1007 hypothetical vents to be used for the simulations. The number of vents in each class is proportional to the probability of activation of the class.

| classi | Volume ( x $10^6$ m$^3$) | | | | |
|---|---|---|---|---|---|
| time (dd) | 0–32 | 32–64 | 64–96 | 96–128 | 128–160 |
| 0 – 15 | 0,24255 | 0,06850 | 0,03418 | | |
| 15 – 30 | 0,11734 | 0,03312 | 0,01648 | 0,01013 | |
| 30 – 60 | 0,11871 | 0,03357 | 0,01663 | 0,01013 | 0,00726 |
| 60 – 90 | 0,06911 | 0,01951 | 0,00968 | 0,00590 | 0,00423 |
| 90 – 120 | 0,04098 | 0,01149 | 0,00348 | 0,00257 | 0,00257 |
| 120 – 150 | 0,03266 | 0,00922 | 0,00454 | 0,00287 | 0,00197 |
| 150 – 180 | 0,02299 | 0,00635 | 0,00318 | 0,00197 | 0,00136 |
| 180 – 210 | 0,01694 | 0,00484 | 0,00242 | 0,00151 | 0,00106 |
| 210 – 240 | | 0,00393 | 0,00197 | 0,00121 | 0,00091 |

Tab. 4: Probabilità di accadimento delle classi di colate.

Table 1: Table of activations, with associated probabilities.

This purposive sampling method of hypothetical vents is different from the one originally proposed by D'AMBROSIO, RONGO, SPATARO, AVOLIO AND LUPIANO (2006), subsequently refined by CRISCI, AVOLIO, BEHNCKE, D'AMBROSIO, DI GREGORIO, LUPIANO, NERI, RONGO AND SPATARO (2010). Non-uniform grids are commonly used to study particular areas of interest in computational domains: for instance, by this approach, BLOTTNER (1975) tried to capture the turbulence in a boundary layer. While non-uniform grids frequently appear in adaptive methods, they may also be used in a "static" environment, as in BOÑGOLAN, DUAN, FISCHER, OZGOKMEN AND ILIESCU (2007), in which the grid was purposely set finer at the inlet, coarser
downstream, to better capture the dynamics of an evolving gravity current. Additionally, the non-uniform grid turned the zones of varying danger on Mt. Etna into a die, which was used in the simulations.
Lava flows were simulated by using the model SCIARA. In particular, the *fv*2 release of the model was considered (SPATARO, Avolio, Lupiano, Trunfio, Rongo, D'Ambrosio, 2010).

### ASSUMPTIONS ON TIME, PLACE AND TYPE OF ACTIVATIONS

Right now, there is no agreement on when the next activation for Mt. Etna will occur. Lupiano previously calculated return times for events using a Poisson distribution, a common assumption for inter-arrival times for events that are considered 'random'. This was implemented for this set of experiments by designing a Poisson-weighted die to determine the approximate number of events in a century, say. Since we considered 52 flank activations in the past 396 years, we used a Poisson distribution with an assumed mean (and variance) of 13. To simulate 50 years, we assume a mean of seven. The place of occurrence for the average of 13 activations will follow the PDF in Fig. 1, i.e, we expect half of activations to happen in the red areas (where there is the greatest density of vents), a quarter to happen in the orange areas, an eighth in the yellow, and so on, following the logarithmic distribution of probabilities of the PDF. Intuitively, the high-danger areas are favored as places of activation.

Determining the type of activation made use of Lupiano's classification (Table 1) and the calculated probabilities therein, implemented as a 41-sided weighted die, using the associated probabilities as weights. In brief, the most popular type of event (less than 15 days duration, less than 32 x $10^6$ m$^3$ ejecta) is favoured to occur.

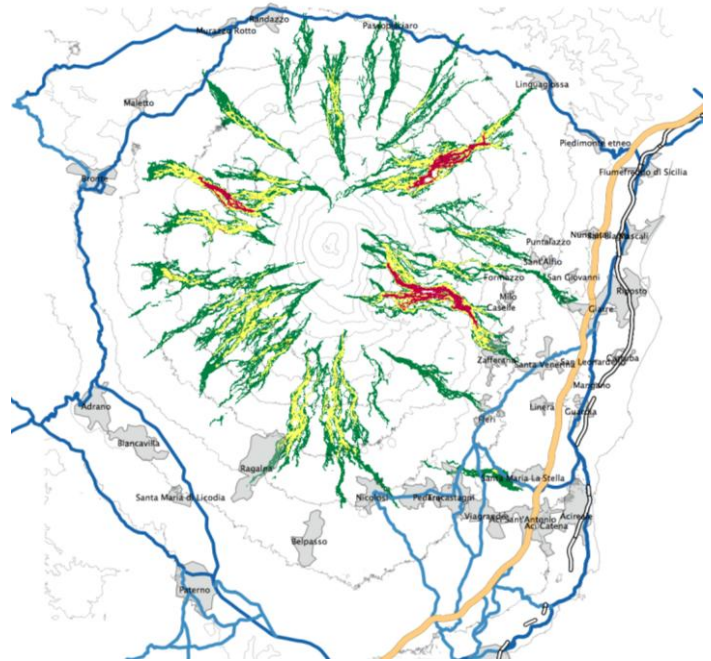

*Fig. 2* – Original experiment on a uniform grid, exhaustive events (all 41 types)

### VALIDATION OF NEW METHODS

Figure 2 shows the 'next activation scenario' from the original experiments by D'AMBROSIO, RONGO, SPATARO, AVOLIO AND LUPIANO (2006): over a uniform grid and an exhaustive testing of events (all 41 types were simulated, so more than 41000 activations were simulated). Figure 3 shows a similar exhaustive experiment, but now on the non-uniform grid of vents. Each figure is 'relatively scaled', ie, red represents the highest risk in each figure, although the calculated (absolute) probability in each scenario may be different. In the original experiment (Fig. 2), the highest absolute probability was 0.09, in the non-uniform grid experiment, (Fig. 3), it was 0.066. The earlier experiments over a uniform grid of vents appear to over-state the danger from the low-danger zones in the PDF, Fig. 1. This might explain the higher calculated probability, compared to Fig. 3, which had a more refined topography in the high-danger zones. All displays in this paper follow a logarithmic scaling of colors, ie, red shows the highest danger, yellow shows half the danger, green a fourth, and blue an eighth. Comparing the two maps shows a very good general agreement, but we note that the non-uniform grid gives higher invasion risks for Nicolosi (southeast) and Ragalna (south). This suggests that the threat to these towns from the next activation on Etna are coming from the high-danger zones (shown as red) in the PDF (Fig. 1), since these have a more refined topography. Figure 4 shows the results of the probabilistic approach; only about 1000 activations were simulated. It is also relatively scaled, with a maximum absolute probability of 0.09. Here, dice were thrown to determine the location of the vent that will be allowed to activate (recall red or high-danger zones in Fig. 1 are favoured); and finally, the 41-sided die is thrown to determine the type of activation for the allowed vent (recall this die favours the most popular type of event, short duration, small volume ejected). We note a good general agreement, except in the direction of Bronte (west). Since we expect agreement in the limit of the probabilistic experiment, we surmise that the invasion risk in the direction of Bronte is coming from the more 'rare' events, possibly originating from the less dangerous zones in Fig. 1.

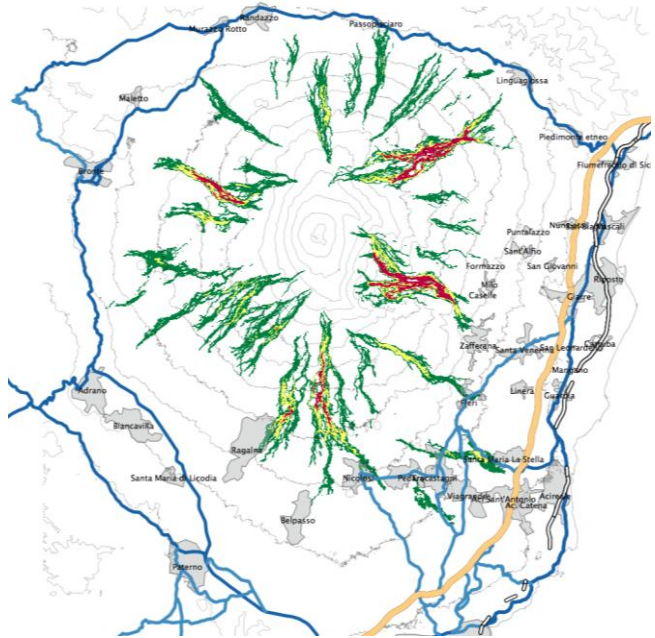

*Fig. 3* – Non-uniform grid, exhaustive events.

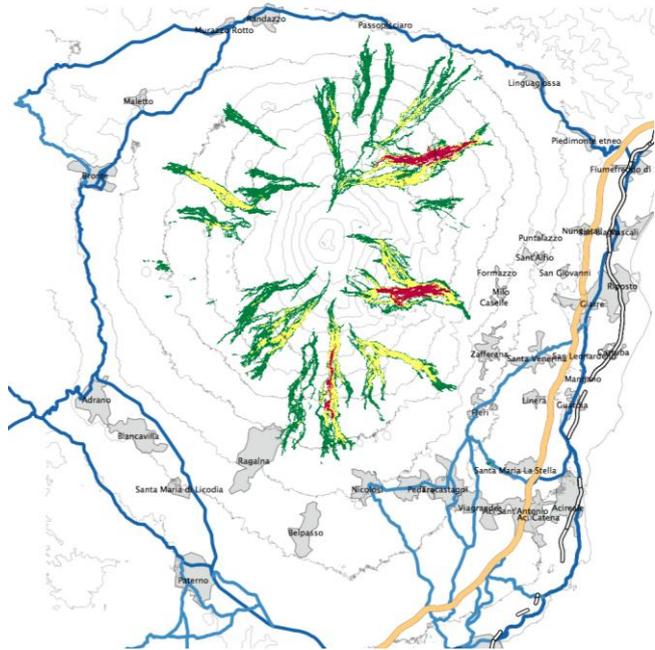

*Fig. 4* – Non-uniform grid, probabilistic events.

## SIMULATING A CENTURY

Experiments were run on the non-uniform grid of 1007 vents to simulate the next 100 years. Averaging the volcano's behaviour for the past 400 years, on average, only 13 of these vents will activate in a century, and the distribution is Poisson.

**For each run**

- ❖ [ Reset to 2013 topography
- ❖ [ We throw the Poisson-weighted die to determine the number of activations **N** to be allowed. On average, 13 will be allowed.
- ❖ [ In random order, the **N** vents will:
  - ( Throw the 41-sided weighted die, to determine the type of activation. See Table 1.
  
  ( Each vent erupts in turn. Each activation will change the topography, and the resulting topography is preserved between activations.
- ❖ [ End Run
- ❖ [ Store final topography for this run.

We did 78 repeats of this experiment, and averaged the resulting probabilities and topographies. This will show us the most likely invasion risk and topography of Mt. Etna in the next 100 years.
Predicting the scenario for the next 50 years proceeds similarly, but only seven vents were allowed to activate, on average. For 25 years, only three vents were allowed to activate, on average. In the case of the next activation, only one vent was allowed to erupt in each run. The advantage over the previous, deterministic approach, is that simulations were faster. It only takes a week or less to do all the simulations properly evaluated and prepared for.

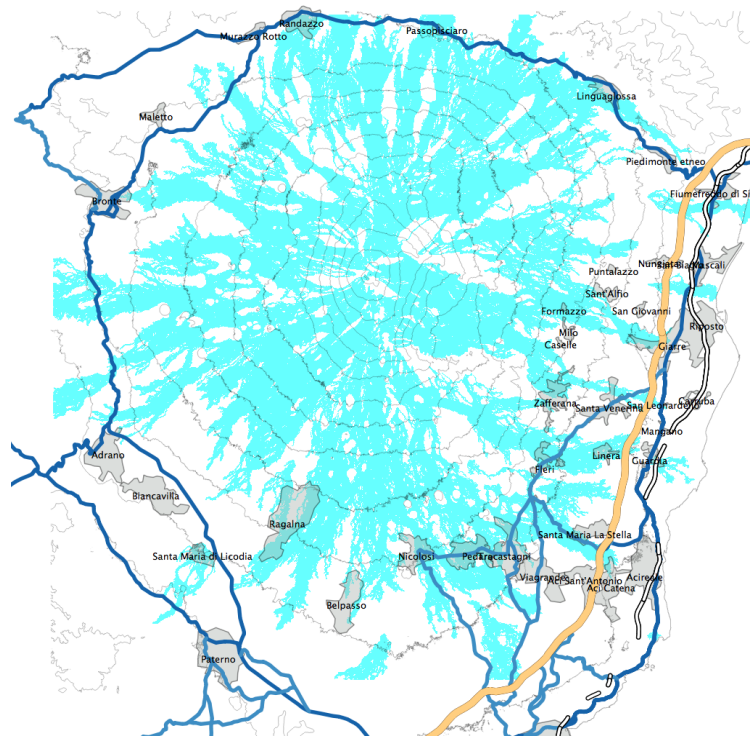

*Fig. 5* – Expected topographic change on Mt. Etna after the next activation (0-3.125 meters).

## RESULTS AND DISCUSSIONS

**Changes to Topography**

A useful feature of these experiments is that they allow us to track the changes to Etna's topography over long periods, e.g., the next century, and forecast flows and build-ups are in good agreement with what Duncan, Chester and Guest (1981) described as 'topographically unprotected areas' . Figures 5-8 are now scaled **absolutely**, ie, colors are comparable across the maps, and the legend is in Figure 6, in meters. Figure 5 gives the expected 'build-up' after the next activation, (0-3.125 meters); Figure 6 (25 years) identifies the directions of Linguaglossa, Zafferana and Nicolosi as sites of build-up; fifty years shows invasion at the border of Ragalna, and damage to lifelines in the area of Passopiciaro and towards Linguaglossa; and 100 years shows major build-ups towards Linguaglossa and Zafferana, intermittent build-ups towards Nicolosi.
An emerging concern would now be danger of sliding from the accumulating deposits, as they grow very near the population areas. Even if there is a low risk of lava invasion, danger from the deposits mobilising as lahar or debris flows has to be

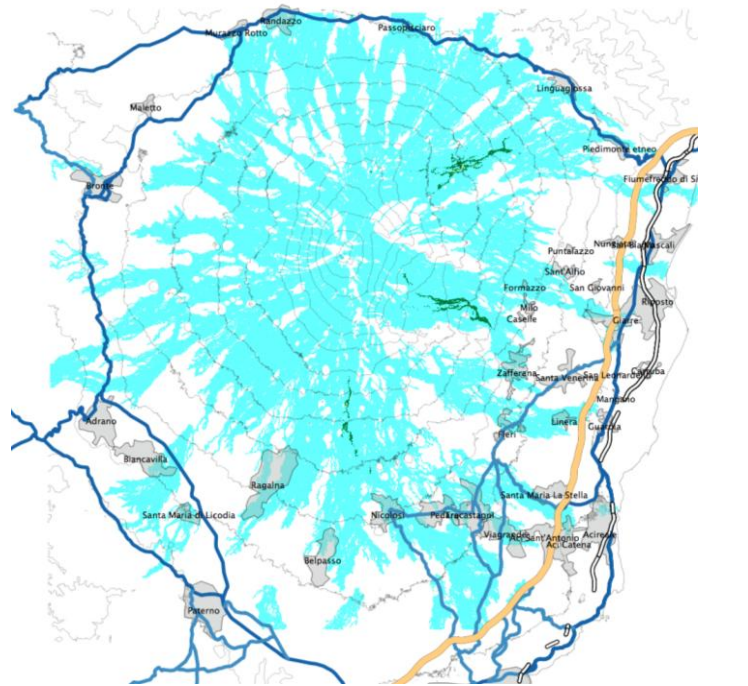

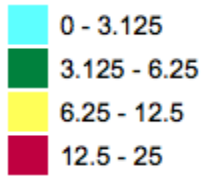

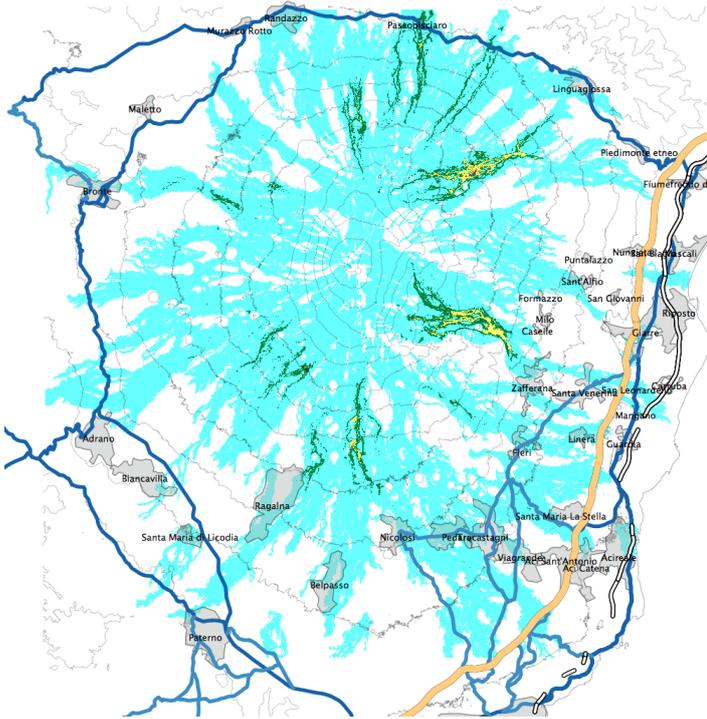

*Fig. 6* – Change to topography in 25 years. Accumulations in the direction of Linguaglossa, Zafferana and Nicolosi.

*Fig. 7* – Fifty-year change to topography; invasion into Ragalna, damage to lifelines in Passopiciaro (north) and area towards Linguaglossa.

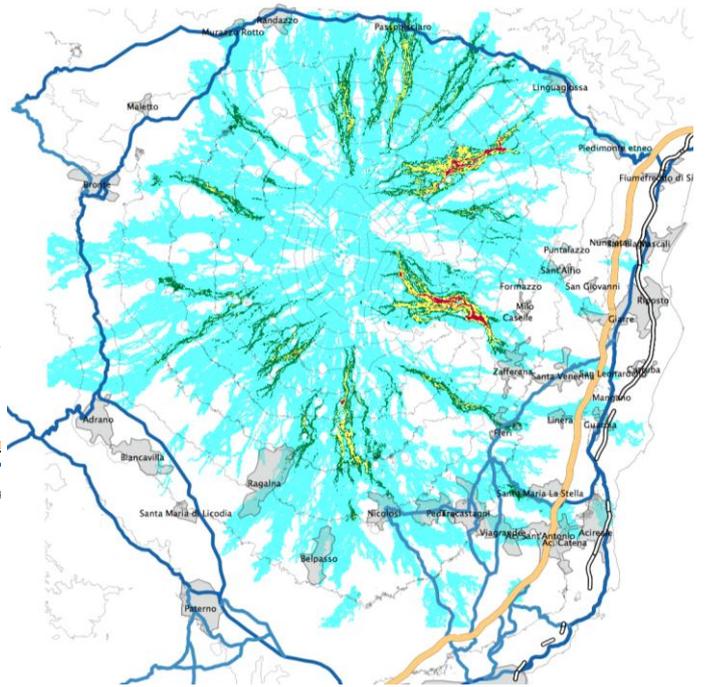

*Fig. 8* – Changes in 100 years; significant build-up towards Linguaglossa, Zafferana, and beginning invasion into Nicolosi.

### Relative Risks of Lava Invasion

Figures 9-12 shows relative risks within each time-frame Maximum calculated (absolute) probabilities are: 0.09 for the next activation (mean 7.76 years); 0.27 for twentyfive years; 0.56 for fifty years, and 0.874999 for one hundred years. Plotting the maximum absolute probabilities and time data, this can be 'perfectly' fit by the cubic polynomial shown below.

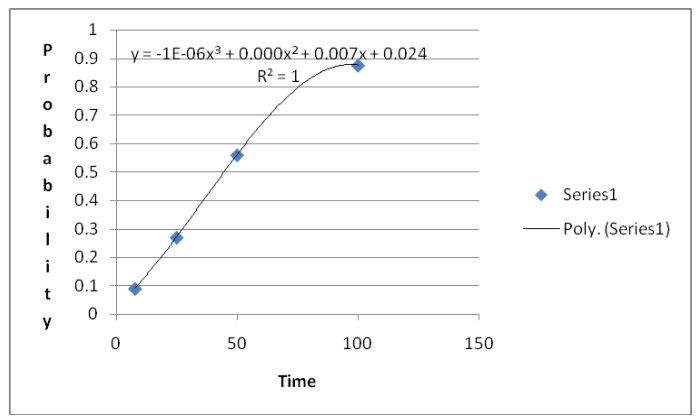

We caution, however, against making interpolations on this. It is to be expected that, with increasing time, the maximum calculated probability should be increasing, asymptotic to one (similar to a 'learning' curve). We show the next four figures only in relative terms.

Each map indicates risks relative to the red areas. Take the 100-year map as an example. We might read it as: If we see lava flowing northeast of Ragalna, then we have a 50 percent chance of lava flowing just outside Ragalna, to the northeast, and a 25 percent chance of lava invasion into its northeast border.

Finally, we 'scaled' the number of repeats for each experiment, to make the series of images comparable. We did approximately 1000 repeats to simulate the next activation, approximately 330 repeats to simulate 25 years, 142 to simulate 50 years, and 78 to simulate 100 years.

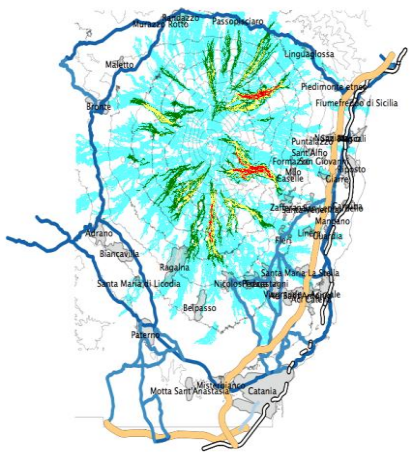

*Fig. 9* Relative Risk for the next activation. Invasion threat for Ragalna; major movement towards Piedmonte, Zafferana and Nicolosi. Threat to lifelines around Passopisciaro.

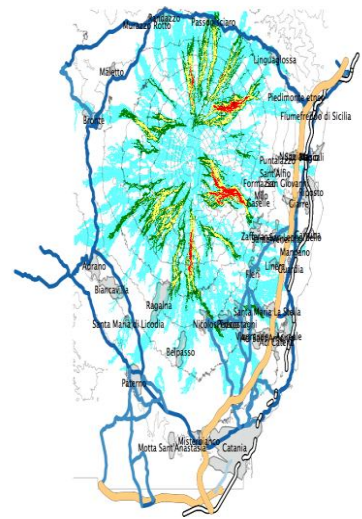

*Fig. 10* Twentyfive Year Relative Risk: Worsening risk of invasion to Ragalna; invasion risk for Nicolosi.

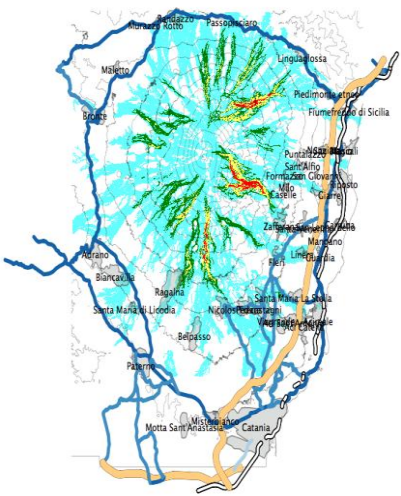

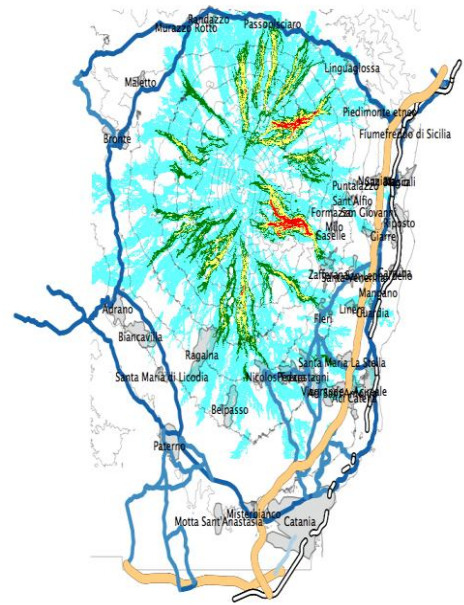

*Fig. 11* Relative Risk for 50 years: Invasion of Nicolosi and threat to lifeline southeast of it; threat getting close to Zafferana, serious threat to lifelines around Passopisciaro.

*Fig. 12* Hundred year Relative Risk: Invasion of Fleri and Zafferana, increasing threats to northeast and southeast, general directions of Linguaglossa, Piedmonte and Zafferana.

**Future Work**

The simulation times are currently being refined to be more detailed, every decade from the present to the next 50 years. With the concerns raised over possible mobilisations of deposits near populated areas, we will also refine the GIS display to show a 1.5 km 'buffer' zone around the populated areas.

**ACKNOWLEDGMENTS**

Boñgolan is currently visiting the University of Calabria and CNR-IRPI with a grant from the Department of Science and Technology, Republic of the Philippines, through the ERDT Consortium. She specially thanks her co-authors for their hospitality.